\documentclass[aps,twocolumn,nofootinbib]{revtex4-2}
\usepackage[latin1]{inputenc}

\begin{document}

\title{Gibbs variational principles and
Boltzmann irreversible theorem}

\author{Mário J. de Oliveira and Silvio R. Salinas}

\affiliation{Universidade de São Paulo,
Instituto de Física,
Rua do Matão, 1371, 05508-090
São Paulo, SP, Brasil}

\begin{abstract}

We analyze the Gibbs variational principles associated with the probability
distributions of (i) an isolated system and (ii) a system at constant
temperature. We give an example of using the Gibbs inequality to obtain the
free energy and analyze the phase diagram of an Ising model with competing
interactions. We also review the Boltzmann irreversible theorem, and show how
it is connected to the Gibbs variational principles. This connection is
established by using the Kolmogorov equation for the evolution of the
probability distribution, which predicts a monotonic increase of entropy for
an isolated system, and a decrease of the free energy for a system in contact
with a thermal reservoir.

\end{abstract}

\maketitle

\section{Introduction}

Variational principles have been a useful resource in the formulation of
physics theories \cite{yourgrau1979}. Variational formulations have even been
considered as more fundamental, as they conform to a certain theological
vision of science, which of course cannot be scientifically validated. In
general, variational principles are equivalent to usual formulations and can
lead us to a deeper understanding of some theories.

In geometric optics, Fermat introduced a minimum principle associated with the
shortest time that it takes for a ray of light to travel between two given
points along a certain path. Consider the path integral between two points,
\begin{equation}
{\cal I}=\int_1^2 \frac{ds}{v},
\end{equation}
where $ds$ is an element of distance and $v$ is the absolute value of the
speed of light. The integral ${\cal I}$ represents the time it takes for
the ray of light to travel along the path between the two given points. Fermat
principle states that the actual path traveled by light corresponds to the
minimum value of this integral.

In mechanics, in the domain of statics, the principle of virtual work of
Johann Bernoulli is a variational principle. It was extended by d'Alembert to
dynamics. On the basis of this principle, Lagrange formulated analytical
mechanics \cite{oliveira2021}. Lagrange considers the time integral
\begin{equation}
{\cal I}=\int_1^2{\cal L}dt,
\end{equation}
between two given instants of time, where ${\cal L}=K-V$ is the difference
between the kinetic energy $K$ and the potential energy $V$ of a mechanical
system. Among the various trajectories that we can imagine, the only one that
occurs, in agreement with the Newton equations of motion, corresponds to the
smallest value of ${\cal I}$. The Hamilton formulation of mechanics is
based on a minimum principle involving a function that is written in terms of
the positions and momenta of a mechanical system, which is obtained from
${\cal L}$ by a Legendre transformation.

In statistical mechanics, Gibbs used the thermodynamic potentials to introduce
some variational principles \cite{gibbs1902,brown1964,falk1970}. According to
thermodynamics, the entropy of an isolated system reaches its largest value in
the equilibrium state. If the system is kept at constant temperature, it
reaches its lowest free energy potential in the equilibrium state. Gibbs
variational principles are the translations of these propositions into
statistical mechanics, in which case the states of the systems are understood
as probability distributions and the equilibrium state is given by the
Gibbs distribution.

Another approach to achieving an equilibrium distribution was proposed by
Maxwell \cite{maxwell1860,maxwell1867} and by Boltzmann
\cite{boltzmann1872,cercignani1988,salinas2001,kremer2010}. The historical
``Boltzmann method'' leads to the well-known transport equation for
the molecular probabilities, and to the $H$-theorem for the irreversibility
of the physical processes. The Gibbs proposal to obtain
the equilibrium state from variational principles does not use a dynamic
process; in other words, it does not need the consideration of the time
evolution of the state. However, a dynamic process that leads to the
equilibrium probability was considered by Maxwell and by Boltzmann, who
succeeded in reaching a fundamental transport equation.

We initially discuss the two Gibbs variational principles that lead to the
equilibrium probability distributions of an isolated system and of a system in
contact with a heat reservoir. The Gibbs inequality for the free energy has
been rediscovered and used by several investigators
\cite{peierls1938,girardeau1973}. In the quantum mechanical version,
it is known as the
Bogoliubov inequality. This Gibbs-Bogoliubov inequality has been used in the
1970s to obtain the mean-field behavior of the well-known
Blume-Emery-Griffiths (BEG) model \cite{BEG1971}. A few years later, by the
suggestion of Luiz Guimarães Ferreira, we have taken advantage of the
Gibbs inequality to analyze some temperature and applied field effects in spin
models, which are not obtained in the context of simple mean-field
approximations, but which were detected in experimental results in our
laboratory of low temperature physics \cite{ferreira1977}. Nowadays, the use
of the Gibbs-Bogoliubov inequality to obtain mean-field results is already
incorporated in standard textbooks of statistical mechanics
\cite{Callen1985,chandler1987}.
We then present a brief discussion of our early use of this Gibbs-Bogoliubov
inequality to analyze the thermodynamic behavior of the
axial-next-nearest-neighbor Ising (ANNNI) model \cite{Yokoi1984}, which is
known to display one of the richest phase diagrams of the literature. Besides
establishing the full phase diagram, it has been feasible to use our
variational method to go beyond a first mean-field approximation, and check
the effects of the introduction of some statistical fluctuations
\cite{Tome1987}.

In the final sections, we turn to the analysis of the dynamic behavior.
However, instead of using the Boltzmann approach, we introduce a more general
formalism, which we call Boltzmann-Kolmogorov equation. We point out that the
Boltzmann equation governs the evolution of the one-particle probability
distribution. In contrast, our proposal of a Boltzmann-Kolmogorov equation
refers to the probability associated with all particles of the system.

\section{Gibbs first variational principle}

The Gibbs probability distribution refers to a mechanical system whose states
define a phase space $(q,p)$, where $q$ represents the set of position
coordinates and $p$ is the set of momenta. For an isolated system, the Gibbs
distribution $\rho_{m}$ inphase space $(q,p)$ is given by the microcanonical
form \cite{gibbs1902,salinas2001},
\begin{equation}
\rho_m=\left\{
\begin{array}
[c]{cc}
1/W, & \quad\mathrm{if}\quad E\leq\mathcal{H}\leq E+\Delta E,\\
0,   & \quad\text{otherwise,}
\end{array}
\right.
\label{27}
\end{equation}
where ${\cal H}={\cal H}(q,p)$, with $E\leq\mathcal{H}\leq E+\Delta E$,
is the energy function, and $W$ is a normalization constant,
\begin{equation}
W=\int_{R}dqdp,
\end{equation}
which can be understood as the volume of the region $R$ of phase space defined
by $E\leq\mathcal{H}\leq E+\Delta E$. The entropy is given by the Gibbs
expression,
\begin{equation}
S=-k\int\rho\ln\rho dqdp,
\label{38}
\end{equation}
where $k$ is the Boltzmann constant, and $\rho(q,p)$ is any probability
distribution that vanishes outside the region $R$ of the phase space $(q,p)$.
The function $S$ reaches the largest value when $\rho$ is the microcanonical
Gibbs probability distribution, $\rho_{m}$. The largest value of $S$ is the
actual value $S_{m}$ for the system at equilibrium, which is given by
\begin{equation}
S_{m}=-k\int\rho_{m}\ln\rho_{m}\,dqdp=k\ln W.
\end{equation}
Therefore, we write
\begin{equation}
S_{m}\geq S,
\label{32}
\end{equation}
which is the expression of the Gibbs first variational principle.

We now demonstrate this result according to the arguments of Gibbs in his
influential book on the principles of statistical mechanics \cite{gibbs1902}.
We then introduce a quantity $W'=W'(q,p)$ and
write the density in phase space,
\begin{equation}
\rho=\frac{1}{W'},\qquad E\leq{\cal H}\leq E+\Delta E.
\label{37}
\end{equation}
The normalization of $\rho$ leads to the condition
\begin{equation}
\int_{R}\frac{1}{W'}dqdp=1.
\label{39}
\end{equation}
Therefore, using this notation, we have to show that
\begin{equation}
-\int_{R}\frac{1}{W}\ln\frac{1}{W}
\,dqdp\,\geq\,
-\int_{R}\frac{1}{W^{\prime}}\ln\frac{1}{W'}\,dqdp,
\end{equation}
which can also be written as
\begin{equation}
\ln W\,\geq\,\int_{R}\frac{1}{W'}\ln W'\,dqdp,
\end{equation}
from which we also write
\begin{equation}
\int_{R}\frac{1}{W'}\ln W' dqdp
\leq\int_{R}\frac{1}{W'}\ln W\,dqdp,
\end{equation}
where we have used equation (\ref{39}) and we have taken into account that $W
$ is constant. This inequality can also be written as
\begin{equation}
\int_{R}\frac{1}{W'}\ln\frac{W}{W'}
\,dqdp\geq0,
\end{equation}
since $W$ is a constant related to the constant microcanonical density
$\rho_{m}$. Taking into account that $W$ is a constant, we finally write
\begin{equation}
\int_{R}\left(\frac{1}{W^{\prime}}\ln\frac{W}{W'}
+\frac{1}{W}-\frac{1}{W^{\prime}}\right)dqdp\geq0.
\end{equation}
To prove this final relation, it is sufficient to resort to a well-known
algebraic property of the convex logarithmic functions. In fact, it easy to
see that the positivity of this last equation comes from the inequality
\begin{equation}
a\,\ln a\,\geq\,a-1,\quad\text{for\quad}a=\frac{W}{W'}>0.
\end{equation}

\section{Gibbs second variational principle}

For a system at constant temperature, the Gibbs equilibrium distribution is
called canonical. It is given by \cite{gibbs1902,salinas2001}
\begin{equation}
\rho_{c}=\frac{1}{Z}e^{-{\cal H}/kT},
\label{61a}
\end{equation}
where $Z$ is a normalization constant,
\begin{equation}
Z=\int e^{-{\cal H}/kT}\,dqdp.
\label{61c}
\end{equation}
This distribution corresponds to the maximum value of $S$, given by the Gibbs
formula for the entropy, eq. (\ref{38}), among the probability distributions
such that
\begin{equation}
U=\int{\cal H}\rho\,dqdp
\label{52}
\end{equation}
is equal to a value $E$ given by
\begin{equation}
\int{\cal H}\rho_{c}\,dqdp=E.
\end{equation}

We now use the technique of Lagrange multipliers, which is equivalent to
finding the minimum of $F=U-TS$, where $T$ is understood as a Lagrange
multiplier. The explicit form of $F$ is given by
\begin{equation}
F=\int{\cal H}\rho\,dqdp+kT\int\rho\ln\rho\,dqdp,
\end{equation}
and can be understood as a functional of $\rho$. Denoting by $F_{c}$ the value
of $F$ for which $\rho$ is equal to the canonical distribution $\rho_{c}$, the
second Gibbs variational principle may be written as
\begin{equation}
F_{c}\leq F.
\label{43}
\end{equation}
It should be pointed out that this inequality is valid for any distribution
$\rho$ and not just for the particular case in which the integral (\ref{52})
is equal to $E$. This is a consequence of the introduction of the Lagrange
multiplier, which makes the variations in $\rho$ independent and not
restricted to the condition that the integral (\ref{52}) is equal to $E$.

We remark that $F_{c}$ is given by
\begin{equation}
F_{c}=\int{\cal H}\rho_{c}\,dqdp+kT\int\rho_{c}\ln\rho_{c}\,dqdp.
\end{equation}
Using (\ref{61a}), we obtain
\begin{equation}
F_{c}=-kT\ln Z.
\end{equation}
If we write the density in the form
\begin{equation}
\rho=\frac{1}{Z_{0}}e^{-{\cal H}_{0}/kT},
\label{61b}
\end{equation}
where $F_{0}=-kT\ln Z_{0}$, the function $F$ can be written as
\begin{equation}
F=F_{0}+\langle{\cal H}-{\cal H}_{0}\rangle,
\end{equation}
where the average $\langle\ldots\rangle$ is calculated by using the
distribution $\rho$. The inequality thus can be written in a very convenient
form,
\begin{equation}
F_{c}\leq F_{0}+\langle{\cal H}-{\cal H}_{0}\rangle.
\label{43a}
\end{equation}

The second Gibbs variational principle is usually written in the form of
equation ({\ref{43a}). It has been adapted to quantum statistical mechanics by
Peierls \cite{peierls1938} and by Bogoliubov \cite{girardeau1973}. In both
versions, the principle was used by Ferreira, Salinas and Oliveira
\cite{ferreira1977} for the formulation of variational methods for obtaining
the properties of cooperative statistical models that exhibit phase
transitions and critical phenomena.

To demonstrate the second Gibbs variational principle, we start by writing
$Z$ in the form
\begin{equation}
Z=Z_{0}\int e^{-\beta({\cal H}-{\cal H}_{0})}\rho dqdp,
\end{equation}
which is equivalent to
\begin{equation}
Z=Z_{0}\,\langle e^{-\beta({\cal H}-{\cal H}_{0})}\rangle
\end{equation}
where $\beta=1/kT$, and the average is calculated using the distribution
$\rho$. We then use the algebraic inequality \cite{falk1970}
\begin{equation}
\langle e^{-x}\rangle\geq e^{-\langle x\rangle},
\end{equation}
and write
\begin{equation}
\langle e^{-\beta({\cal H}-{\cal H}_{0})}\rangle\geq e^{-\beta
\langle{\cal H}-{\cal H}_{0}\rangle}.
\end{equation}
If we replace this result in the previous equation, and take the logarithm of
both sides, we finally find that
\begin{equation}
\ln Z\geq\ln Z_{0}-\beta\langle{\cal H}-{\cal H}_{0}\rangle.
\end{equation}
Taking into account that $F_{c}=-kT\ln Z$ and $F_{0}=kT\ln Z_{0}$, we finally
obtain the inequality (\ref{43a}).

\section{Mean-field treatment of the ANNNI model}

We now use the second Gibbs inequality to establish the phase diagram of a
paradigmatic Ising model on a cubic lattice, with ferromagnetic
nearest-neighbor interactions and the addition of competitive
antiferromagnetic second-neighbor interactions along an axial direction, which
has been called Axial-Next-Nearest-Neighbor Ising or ANNNI model.

In its simplest form, the Hamiltonian of this ANNNI model is given by
\[
{\cal H}=-\sum_{x,y,z}
[J_{1}S_{x,y,z} (S_{x+1,y,z}+S_{x,y+1,z})
\]
\begin{equation}
+J_{1}S_{x,y,z}S_{x,y,z+1}+J_{2}S_{x,y,z}S_{x,y,z+2}],
\end{equation}
where $S_{x,y,z}=\pm1$ is an Ising spin on the $(x,y,z)$ site
of a cubic lattice with unit lattice parameter. Interactions between first
neighbors are ferromagnetic ($J_{1}>0$), and interactions between second
neighbors along the $z$ direction favor an antiferromagnetic alignment
($J_{2}<0$). It is known that this axial competition between ferro and
antiferro alignments gives rise to a Lifshitz point and a succession of
modulated phases in a phase diagram in terms of temperature $T$ and the
parameter of competition $p=-J_{2}/J_{1}>0$ \cite{Yokoi1984}.

Mean-field solutions for this problem are based on the Gibbs inequality, given
by equation (\ref{43a}), with the trial Hamiltonian
\begin{equation}
\mathcal{H}_{0}=-\sum_{z}h_{1}(z) =-\sum_{x,y}\sum_{z}\eta
_{z}^{(1)}\,S_{x,y,z}
\label{h0}
\end{equation}
where $\{\eta_{z}^{(1)}\}$ is a set of trial
fields to be determined. The solutions of this mean-field problem have been
published in a pioneer paper by Yokoi, Coutinho-Filho and Salinas
\cite{Yokoi1984}, which also contains a complete analysis of the $T$-$p$ phase
diagram, including connections with other calculations and with a Landau
expansion in the vicinity of the paramagnetic border.

We now use the method proposed by Ferreira, Salinas and Oliveira
\cite{ferreira1977} to go beyond the mean-field results. In the pair
approximation, which corresponds to the first stage of the method, we consider
a certain number of pair interactions in a modified trial Hamiltonian,
\begin{equation}
\mathcal{H}_{0}=-\sum_{z}[\sum_{x,y}h_{1}(z)
+ \sum_{x,y}^{(1)} h_{2}(z)
+ \sum_{x,y}^{(2)} h_{2}'(z)]  ,
\end{equation}
where $h_{1}(z)$ is the usual one-site term, and the
independent two-site terms are given by
\begin{equation}
h_{2}(z) = J_{1}S_{x,y,z}S_{x+1,y,z}+\eta_{z}^{(2)}
[S_{x,y,z}+S_{x+1,y,z}]
\end{equation}
and
\begin{equation}
h_{2}^{\prime}(z)  =J_{1}S_{x,y,z}S_{x,y+1,z}
+\eta_{z}^{(2)}\,[S_{x,y,z}+S_{x,y+1,z}],
\end{equation}
with the interactions restricted to a certain number of nearest-neighbor sites
belonging to the same $x$-$y$ plane.

Now it is straightforward to write
\[
G_{0}=-kTn_{1}\sum_{z}
\ln[2\cosh\beta\eta_{z}^{(1)}]
\]
\begin{equation}
-kT n_2 \sum_{z}
\ln[2 e^{\beta J_0}
\cosh\beta\eta_{z}^{(2)}+2 e^{-\beta J_0}]  ,
\end{equation}
where $n_{1}=N^{2}-2n_{2}$ is the number of independent sites, $n_{2}$ is the
number of isolated pairs, and $N^{2}$ is the number of independent sites in
each $x,y$ plane. We also write
\[
\langle{\cal H} - {\cal H}_{0}\rangle
=- (2N^{2}-n_{2}) J_0 \sum_z m_z^2
\]
\[
-\frac{1}{2}N^{2}J_{1} \sum_z m_{z}(m_{z-1}+m_{z+1})
\]
\[
-\frac{1}{2}N^{2}J_{21} \sum_z m_z (m_{z-2}+m_{z+2})
\]
\begin{equation}
+ \sum_z
[N^{2}\eta_{z}^{(1)}+2n_{2}(\eta_{z}^{(2)}-\eta_{z}^{(1)})]  m_{z}.
\end{equation}

From the minimization of this trial free energy, it is straightforward to
write equations of state and discuss all of the delicate features of the phase
diagrams. At this point the reader is invited to check the detailed results in
the paper by Tom\'{e} and Salinas \cite{Tome1987}. However, there is a
delicate question about the choice of the number $n_{2}$ of isolated pairs.
According to our previous work, Tom\'{e} and Salinas resort to the first few
and relatively easy to obtain terms of the exact high-temperature expansion of
the free energy. According to this choice, we should take $n_{2}=2N^{2}$,
which corresponds to an analytic continuation of our expressions, and which
considerably simplifies the equations. Incidentally, this kind of choice also
leads to the well known Bethe-Peierls equation of state for the simple
Ising ferromagnet.

In conclusion, according to Tom\'{e} and Salinas \cite{Tome1987}, the general
features of the variational layer-by-layer mean-field phase diagram of the
ANNNI model are not changed by the inclusion of some pair spin interactions.
The paramagnetic lines are depressed, in agreement with analyses on the basis
of high-temperature series expansions. In the phase diagram in terms of $p$
and $T$, the paramagnetic-modulated and the paramagnetic-ferromagnetic
critical lines still meet smoothly at a Lifshitz point with the first-order
ferromagnetic-modulated transition line. The more refined mean-field
calculations show that the Lifshitz point, as in the experiments on the
magnetic compound MnP, is and inflection point of the paramagnetic border.
Numerical calculations show that the main commensurate phase are stable under
the introduction of some spin fluctuations.

\section{Liouville Equation}

The search for an evolution equation that leads to the final equilibrium state
of a system was carried out by Boltzmann within the context of the classical
kinetic theory of gases. Boltzmann proposed a nonlinear equation that governs
the time evolution of the one-particle probability distribution.

Let us call $x_{i}$ the collection of the Cartesian components of the position
and the momentum of particle $i$. The probability distribution $\rho(x)$ is
then a function of all variables $x_{i}$ which we are denoting by $x$. The
one-particle distribution function $f_{i}(x_{i})$ considered by Boltzmann can
be understood as the marginal distribution given by
\begin{equation}
f_{i}(x_{i})=\int\rho(x)\prod_{j(\neq i)}dx_{j},
\end{equation}
where the integration is over all variables $x_{j}$ except $x_{i}$. One of the
most relevant results of Boltzmann refers to the behavior of a function
$S_{B}$, given by
\begin{equation}
S_{B}=-k\int f_{i}(x_{i})\ln f_{i}(x_{i})dx_{i},
\end{equation}
which has been shown to be an increasing function of time and to reach its
maximum value at equilibrium. This has been known as the $H$-theorem. As
$S_{B}$ refers to the distribution of one particle, we are tempted to say that
the entropy of the system is $NS_{B}$ where $N$ is the number of particles.
However, for a system consisting of interacting particles, the entropy $S$
defined by the Gibbs formula (\ref{38}) cannot be $NS_{B}$.

An alternative to the Boltzmann equation is the Liouville equation
\cite{salinas2001},
\begin{equation}
\frac{\partial\rho}{\partial t}=\{{\cal H},\rho\}
\label{71}
\end{equation}
where the right-hand side are the Poisson brackets,
\begin{equation}
\{\mathcal{H},\rho\}=\sum_{i}(\frac{\partial{\cal H}}{\partial q_{i}}
\frac{\partial\rho}{\partial p_{i}}
-\frac{\partial{\cal H}}{\partial p_{i}}
\frac{\partial\rho}{\partial q_{i}}),
\end{equation}
which govern the time evolution of $\rho$. The Liouville equation is
understood as describing an isolated system because the energy function
${\cal H}(q,p)$ is strictly conserved along a path in phase space. However,
using the Liouville equation one can show that the entropy $S$ given by the
Gibbs formula (\ref{38}) does not increase in time. In fact, it is constant in
time. It is easy to show this result by writing the Gibbs entropy,
\begin{equation}
S=-k\int\rho\ln\rho dx,
\label{28}
\end{equation}
and taking the derivative with respect to time,
\begin{equation}
\frac{dS}{dt}=-k\int\frac{\partial\rho(x)}{\partial t}\ln\rho(x)dx.
\label{28a}
\end{equation}
It should be remarked that there is a second term, but it vanishes because
$\rho$ is normalized. Using the Liouville equation, we find
\begin{equation}
\frac{dS}{dt}=-k\int\{{\cal H},\rho\}\ln\rho(x)dx.
\label{28b}
\end{equation}
If we integrate by parts, it is not difficult to show that
\begin{equation}
\frac{dS}{dt}=0.
\end{equation}
Therefore, in this formulation $S$ is constant in time. We then have to
abandon the Liouville equation and try to look for another formalism that
should be capable of predicting an increase of entropy.

\section{Boltzmann-Kolmogorov Equation}

The assumptions contained in the Boltzmann reasoning to reach his transport
equation can be retrospectively understood if we consider a Markov stochastic
process \cite{oliveira2019,oliveira2024}. The Boltzmann assumption was called
collision number hypothesis (\textit{Stosszahlansatz}) by the Ehrenfests, in
their famous review of the Boltzmann method \cite{ehrenfest1907}. The crucial
point of the Boltzmann derivation is the introduction of the conditional
probability that two colliding particles have certain velocities given the
velocities of the particles at an earlier time. Referring to two particles $i$
and $j$, this quantity is proportional to the rate $w_{ij}(x_{i}^{\prime
},x_{j}^{\prime};x_{i},x_{j})$ of the transition $(x_{i},x_{j})\rightarrow
(x_{i}^{\prime},x_{j}^{\prime})$ connecting the states of two particles. The
total rate $w(x^{\prime},x)$ of a transition $x\rightarrow x^{\prime}$ is the
sum of the transition rates associated with two particles, that is,
\begin{equation}
w(x^{\prime},x)
=\sum_{ij}w_{ij}(x_{i}^{\prime},x_{j}^{\prime};x_{i},x_{j}).
\label{78}
\end{equation}
Given two states $x$ and $x^{\prime}$ we assume that just one of the several
terms of the right-hand side of this equation is nonzero.

Instead of proceeding according to Boltzmann, we may use the transition rates
to establish a Kolmogorov equation \cite{kolmogorov1931,gardiner2009}, which
has been proposed to govern the evolution of the probability distributions of
a Markovian stochastic process. We then write
\begin{equation}
\frac{\partial\rho}{\partial t}=\int[w(x,x^{\prime})\rho(x^{\prime
})-w(x^{\prime},x)\rho]dx^{\prime}.
\label{73}
\end{equation}
However, this equation includes only transitions due to a stochastic force. To
be more precise, it is restricted to forces that we assume to \textit{be
stochastic, as in the }collisions of two hard spheres. To include
deterministic transitions coming from conservative forces, we should add the
Poisson term of the Liouville equation. The evolution equation is then written
as
\begin{equation}
\frac{\partial\rho}{\partial t}=\{\mathcal{H},\rho\}
+\int[w(x,x^{\prime})\rho(x^{\prime})-w(x^{\prime},x)\rho(x)]dx^{\prime}.
\label{74}
\end{equation}

Boltzmann did not obtain either equation (\ref{74}) or equation (\ref{73}).
He obtained an equation which can retrospectively be understood as an
approximation to equation (\ref{74}), which consists in writing $\rho$ as a
product of the probability distribution associated with each particle. This
assumption, which is implicit in Boltzmann writings, was called hypothesis of
molecular disorder (\textit{molekularen Unordnung}) in the well-known review
of the Ehrenfests \cite{ehrenfest1907}.

To go beyond the Boltzmann proposal, we keep using the essential properties of
the transition rates which were already used by Boltzmann and by Maxwell. The
rates $w(x^{\prime},x)$ are in accordance with the conservation of energy and
momentum, that is, $w(x^{\prime},x)$ is nonzero only when the energy
$\mathcal{H}(x^{\prime})$ is equal to $\mathcal{H}(x)$. The momentum
associated with state $x^{\prime}$ is equal to the momentum of state $x$.
Another essential property that is crucial for the development of the present
approach is the reversal property, $w(x,x^{\prime})=w(x^{\prime},x) $. This
last property leads to the following form of Boltzmann-Kolmogorov equation,
\begin{equation}
\frac{\partial\rho}{\partial t}=\{\mathcal{H},\rho\}
+\int w(x,x^{\prime})[\rho(x^{\prime})-\rho(x)]dx^{\prime}.
\label{74a}
\end{equation}
We remark that in 1956 Kac \cite{MKac1956} introduced a similar equation,
without the Poisson term, and using transition rates restricted to the
original Boltzmann proposal.

From the property of energy conservation we conclude that the trajectory of a
representative point in phase space obeys the relation $\mathcal{H}(x)=E$,
which is a constant for all points $x$ of this trajectory. Therefore,
$\rho(x)$ will be nonzero only when $\mathcal{H}(x)=E$. Taking into account
that in the stationary state $\rho(x)$ does not depend on time, we conclude
that the probability distribution at the stationary state is the
microcanonical Gibbs distribution, given by (\ref{27}).

We now determine the time variation of the entropy. If we replace equation
(\ref{74a}) in equation (\ref{28a}), we have
\begin{equation}
\frac{dS}{dt}=-k\int w(x,x^{\prime})[\rho(x^{\prime})-\rho(x)]\ln
\rho(x)dxdx^{\prime}.
\end{equation}
As we have shown, the term associated with the Poisson brackets vanishes. We
then write this expression in the equivalent form
\begin{equation}
\frac{dS}{dt}=\frac{k}{2}\int w(x,x^{\prime})[\rho(x^{\prime})-\rho
(x)]\ln\frac{\rho(x^{\prime})}{\rho(x)}dxdx^{\prime}.
\end{equation}
Using the inequality $(a-b)\ln(a/b)\geq0$, we conclude that the integrand is
nonnegative because $w(x,x^{\prime})\geq0$. Therefore,
\begin{equation}
\frac{dS}{dt}\geq0
\label{90}
\end{equation}
which we call the first Boltzmann irreversible theorem.

If $S_{v}$ is the entropy associated with an initial distribution $\rho_{v}$
and if $S_{m}$ is the entropy of the equilibrium state, which is the
microcanonical Gibbs distribution $\rho_{m}$, then from (\ref{90}) it follows
that $S_{m}\geq S_{v}$. We remark that this is the representation of the first
Gibbs variational principle, given by equation (\ref{32}).

\section{Contact with a Thermal Reservoir}

The transition rates that we have considered in the last section describe the
stochastic interaction of molecules. They are in agreement with the
conservation of energy and momentum in addition to the reversal property,
which from now on we denote by $w^{b}(x^{\prime}x)$. The conservation of
energy of the transition rate $w^{b}(x^{\prime}x)$ allows us to say that the
system described by the Boltzmann-Kolmogorov equation (\ref{74}) is an
isolated system. Now we wish to describe an open system in contact with a heat
reservoir at a temperature $T$. In this case, at long times, the system will
be found in the equilibrium state with the Gibbs canonical probability distribution.

The transition rate $w^{c}(x^{\prime},x)$ that describes the contact with the
heat reservoir is assumed to obey the property
\begin{equation}
\frac{w^{c}(x^{\prime},x)}{w^{c}(x,x^{\prime})}=\frac{\exp\left(
-\mathcal{H}(x^{\prime})/kT\right)}
{\exp\left(-\mathcal{H}(x)/kT\right)}.
\label{62}
\end{equation}
The Boltzmann-Kolmogorov equation has now two stochastic terms corresponding
to the internal stochastic interactions described by the rate
$w^{b}(x^{\prime}x)$ and the stochastic interaction due to the
contact with the heat reservoir described by the rate $w^{c}(x^{\prime},x)$.
We then write
\[
\frac{\partial\rho}{\partial t}=\{\mathcal{H},\rho\}
+\int w^{\mathrm{b}}(x,x^{\prime})
[\rho(x^{\prime})-\rho(x)]dx^{\prime}+
\]
\begin{equation}
+\int[w^{c}(x,x^{\prime})\rho\left(  x^{\prime}\right)
-w^{c}(x^{\prime},x)\rho(x)]dx^{\prime}.
\label{61}
\end{equation}

We remark that this equation, without the first integral, was introduced by
Bergmann and Lebowitz \cite{bergmann1955} in 1955 to study nonequilibrium
processes. They used transition rates with the property (\ref{62}) and also
considered the case where the system is in contact with several reservoirs at
distinct temperatures. However, at this point we wish to make connections with
the Gibbs canonical distribution, which occurs when there is just one reservoir.

It is straightforward to show by a simple substitution that the Gibbs
canonical distribution (\ref{61a}) is the equilibrium distribution of equation
(\ref{61}). Initially, we observe that the Poisson brackets vanishes because
$\rho_{\mathrm{c}}$ is a function of $\mathcal{H}$. To see that the second
integral in (\ref{61}) vanishes, it is sufficient to write (\ref{62}) as
\begin{equation}
w^{\mathrm{c}}(x^{\prime},x)\rho_{\mathrm{c}}(x)
=w^{\mathrm{c}}(x,x^{\prime})\rho_{\mathrm{c}}(x^{\prime}).
\end{equation}
Now, taking into account that $w^{\mathrm{b}}(x,x^{\prime})$ vanishes when
$\mathcal{H}(x^{\prime})$ is not equal to $\mathcal{H}(x)$,
we obtain the identity
\begin{equation}
w^{\mathrm{b}}(x^{\prime},x)[\rho_{\mathrm{c}}(x^{\prime})
-\rho_{\mathrm{c}}(x)]=0.
\end{equation}
Using this result we see that the first integral in (\ref{61}) vanishes.

In the previous section, we have seen that the energy was strictly conserved.
In the present case, this no longer happens due to the contact with the
reservoir with which the system exchanges energy. In fact, the energy
exchanged is called heat because we are not considering external work done on
the system. Let $U=\langle H(x)\rangle$ be the average energy,
\begin{equation}
U=\int H(x)\rho(x)dx.
\end{equation}
Deriving this equation with respect to time we have
\begin{equation}
\frac{dU}{dt}=\int H(x)\frac{\partial\rho(x)}{\partial t}dx
\end{equation}
Using the evolution equation (\ref{61}), we obtain
\begin{equation}
\frac{dU}{dt}=\int\left[  H\left(  x^{\prime}\right)  -H\left(  x\right)
\right]  w^{c}(x^{\prime},x)\rho(x)]\,dx^{\prime}dx,
\label{76}
\end{equation}
and we remark that the terms associated with the Poisson brackets and with the
internal stochastic processes vanish. Using the property (\ref{62}), we may
write
\begin{equation}
\frac{dU}{dt} =
-kT\int w^{\mathrm{c}}(x^{\prime},x)\rho(x)
\ln\frac{w^{\mathrm{c}}(x^{\prime},x)}{w^{\mathrm{c}}
(x,x^{\prime})}dx^{\prime}dx.
\label{46a}
\end{equation}
This expression means that the change in the energy of the system is entirely
due to the contact with the heat reservoirs, as it should be expected.

Let us determine the time variation of the entropy. Replacing (\ref{61}) in
the expression (\ref{28a}) we have
\[
\frac{dS}{dt}=k\int w^{\mathrm{b}}(x^{\prime},x)\rho(x)
\ln\frac{\rho(x)}{\rho(x^{\prime})}dxdx^{\prime}
\]
\begin{equation}
+k\int w^{\mathrm{c}}(x^{\prime},x)\rho(x)
\ln\frac{\rho(x)}{\rho(x^{\prime})}dxdx^{\prime},
\label{46b}
\end{equation}
and we remark that the term containing the Poisson brackets vanishes.

We now define the free energy $F$ by $F=U-TS$, and consider the time
derivative of $F$,
\begin{equation}
\frac{dF}{dt}=\frac{dU}{dt}-T\frac{dS}{dt}.
\label{46c}
\end{equation}
We demonstrate that
\begin{equation}
\frac{dF}{dt}\leq0,
\label{66}
\end{equation}
which we call the second Boltzmann irreversible theorem. This result was also
demonstrated by Bergmann and Lebowitz in his paper of 1955 using the equation
they introduced \cite{bergmann1955}, which corresponds to equation (\ref{61})
without the first integral.

Replacing (\ref{46a}) and (\ref{46b}) in (\ref{46c}), we find
\[
\frac{dF}{dt}=-kT\int w^{\mathrm{b}}(x^{\prime},x)\rho(x)\ln\frac{\rho
(x)}{\rho(x^{\prime})}dxdx^{\prime}
\]
\begin{equation}
-kT\int w^{\mathrm{c}}(x^{\prime},x)\rho(x)
\ln\frac{w^{\mathrm{c}}(x^{\prime},x)\rho(x)}
{w^{\mathrm{c}}(x,x^{\prime})\rho(x^{\prime})}dx^{\prime}dx,
\label{49}
\end{equation}
which can be written as
\[
\frac{dF}{dt}=-\frac{kT}{2}\int w^{\mathrm{b}}(x^{\prime},x)
[\rho(x)-\rho(x^{\prime})]\ln\frac{\rho(x)}{\rho(x^{\prime})}
dxdx^{\prime}
\]
\begin{equation}
-\frac{kT}{2}\int[w^{\mathrm{c}}(x^{\prime\mathrm{c}}
(x,x^{\prime})\rho(x^{\prime})]\ln\frac{w^{\mathrm{c}}(x^{\prime},x)
\rho(x)}{w^{\mathrm{c}}(x,x^{\prime})\rho(x^{\prime})}dx^{\prime}dx.
\end{equation}
Using again the inequality $(a-b)\ln(a/b)\geq0$ we see that both integrals are
nonnegative, and we conclude that the right hand side of this equation is
negative or zero, leading us to the inequality (\ref{66}). We remark that
$dF/dt$ vanishes when $\rho$ equals $\rho_{c}$, that is, in equilibrium.
Therefore, $F$ decreases monotonically in time and reaches its minimum values
at equilibrium. Denoting by $F_{c}$ the value of $F$ when $\rho$ equals
$\rho_{c}$, then $F_{c}\leq F$, which is the expression of the second Gibbs
variational principle, given by\ equation (\ref{43}).

\section{Conclusions}

We have shown that the Gibbs variational principles are direct related to the
Boltzmann irreversible theorems. They involve a thermodynamic potential, which
is the entropy $S$ in the case of an isolated system. According to the
Boltzmann theorem, $S$ increases monotonically. Since $S$ is bounded by the
Gibbs principle, then $S$ reaches the equilibrium value in the long run. In
the case of a system at constant temperature, the associated thermodynamic
potential is the free energy $F$. In this case, the Boltzmann theorem states
that $F$ decreases monotonically; since $F$ is bounded by the Gibbs principle,
then $F$ reaches the equilibrium value in the long run.

If we define $\psi=F-F_{c}$ then the second Gibbs principle and the Boltzmann
theorem are written as
\begin{equation}
\Psi\leq0,\qquad\frac{d\Psi}{dt}\leq0.
\label{69}
\end{equation}
and the explicit form of $\psi$ is given by
\begin{equation}
\Psi=kT\int\rho\ln\frac{\rho}{\rho_{c}}dx\,.
\end{equation}
The inequalities (\ref{69}) are the properties that define a Lyapunov
function. In accordance with the Lyapunov theorem, the stability of the
stationary solution of a set of ordinary equations of first order in time is
proved if we construct a function that obeys the inequalities of equation
(\ref{69}). In this case, the set of equations is the Boltzmann-Kolmogorov
equation and the Lyapunov theorem is nothing more than the combination of
Gibbs variational principle and Boltzmann irreversible theorem.

\section*{Acknowledgements}

This paper is dedicated to Luiz Guimar\~{a}es Ferreira, who extensively used
variational methods, including applications to statistical mechanics in
collaboration with the authors \cite{ferreira1977}.

\end{document}